# Magnetic Resonance Spectroscopy Quantification Aided by Deep Estimations of Imperfection Factors and Macromolecular Signal

Dicheng Chen, Meijin, Lin, Huiting Liu, Jiayu Li, Yirong Zhou, Taishan Kang, Liangjie Lin, Zhigang Wu, Jiazheng Wang, Jing Li, Jianzhong Lin, Xi Chen, Di Guo, and Xiaobo Qu, *Senior Member, IEEE*

*Abstract*—*Objective:* Magnetic Resonance Spectroscopy (MRS) is an important technique for biomedical detection. However, it is challenging to accurately quantify metabolites with proton MRS due to serious overlaps of metabolite signals, imperfections because of non-ideal acquisition conditions, and interference with strong background signals mainly from macromolecules. The most popular method, LCModel, adopts complicated non-linear least square to quantify metabolites and addresses these problems by designing empirical priors such as basis-sets, imperfection factors. However, when the signal-to-noise ratio of MRS signal is low, the solution may have large deviation. *Methods:* Linear Least Squares (LLS) is integrated with deep learning to reduce the complexity of solving this overall quantification. First, a neural network is designed to explicitly predict the imperfection factors and the overall signal from macromolecules. Then, metabolite quantification is solved analytically with the introduced LLS. In our Quantification Network (QNet), LLS takes part in the backpropagation of network training, which allows the feedback of the quantification error into metabolite spectrum estimation. This scheme greatly improves the generalization to metabolite concentrations unseen for training compared to the end-to-end deep learning method. *Results:* Experiments show that compared with LCModel, the proposed QNet, has smaller quantification errors for simulated data, and presents more stable quantification for 20 healthy *in vivo* data at a wide range of signal-to-noise ratio. QNet also outperforms other end-to-end deep learning methods. *Conclusion:* This study provides an intelligent, reliable and robust MRS quantification. *Significance:* QNet is the first LLS quantification aided by deep learning.

## I. Introduction

MAGNETIC Resonance Spectroscopy (MRS) is a technique for non-invasive detection of biochemical metabolites in living tissues. It has been clinically used to identify changes in metabolite levels in regions of interest (e.g., brain [1], spinal cord [2], liver [3] and other organs [4]) of *in vivo*. Thus, MRS has great scientific and clinical value for the diagnosis, and therapeutic follow-up of genetic, oncological and degenerative diseases. However, the quantification of metabolites in proton MRS ($^1$H-MRS) is rather complicated. This is because of serious overlaps of metabolite signals, signal distortions due to non-ideal acquisition conditions and interference with background signal that are mostly contributed by MacroMolecules (MMs). How to handle the above problems will be briefly discussed below.

Highly overlapped signals of metabolites are due to the narrow spectral dispersion caused by the limited magnetic field strength of a clinical magnetic resonance scanner. To improve the quantification, a direct way is introducing empirical prior of each metabolite signal. As a most recognized quantification method, LCModel [5] designed a basis-set of metabolic profiles to calculate the contribution of individual metabolites to an observed spectrum of multiple metabolites. The basis-set was acquired from a high-field scanner through time-consuming *in vitro* experiments and may introduce experimental errors (e.g. effects of temperature, pH, etc.) [6], [7]. These errors need correction through complicated processing to adapt the basis-set to *in vivo* human spectrum [6]. In another way, without any real scanning, the quantification errors [8], [9] could be reduced by using the simulated basis-set [10]. Through the Quantum Mechanical and Exponential Model (QMEM), each metabolite

This work was supported by the National Natural Science Foundation of China (62122064, 61971361, 62331021, 62371410), the Natural Science Foundation of Fujian Province of China (2023J02005 and 2021J011184), the President Fund of Xiamen University (20720220063), and Nanqiang Outstanding Talent Program of Xiamen University.

D. Chen, H. Liu, Y. Zhou, Jiayu Li and X. Qu (corresponding author email: quxiaobo@xmu.edu.cn) are with Department of Electronic Science, Fujian Provincial Key Laboratory of Plasma and Magnetic Resonance, Xiamen University, Xiamen 361104, China.
M. Lin is with Department of Applied Marine Physics and Engineering, and Fujian Provincial Key Laboratory of Plasma and Magnetic Resonance, Xiamen University, Xiamen 361104, China.
L. Lin, Z. Wu, and J. Wang are with Philips, Beijing 100016, China.
J. Li is with Shanghai Electric Group CO., LTD, Shanghai 200001, China and was with Xingaoyi Medical Equipment Company, Yuyao 315400, China.
T. Kang and J. Lin are with Department of Radiology, Zhongshan Hospital affiliated to Xiamen University, Xiamen 361004, China.
X. Chen is with the McLean Hospital, Harvard Medical School, Belmont, MA 02478, USA.
D. Guo is with School of Computer and Information Engineering, Xiamen University of Technology, Xiamen 361024, China.
D. Chen and M. Lin are equally contributed to this work.

signal can be generated according to the actual pulse sequence parameters, the *in vivo* environments (e.g., MM or lipid signals) and Imperfection Factors (IFs) of non-ideal acquisition conditions (e.g., deviations of phase, frequency, linewidth, etc.). Current QMEM MRS tools include FID-A [11], GAMMA [10], NMR-SCOPE [12], etc.

IFs of non-ideal acquisition conditions are commonly mathematically modelled using nonlinear factors [13]. For MRS, three main nonlinear factors are commonly considered. 1) Phase shift due to the dominating zero-order phase misadjustment from the phase difference between the excitation and the detection [14]; 2) Frequency shift due to scanner frequency shift, magnetic field inhomogeneity, tissue heterogeneity and variation, subject movement or physiological motion, etc. [15]; 3) Linewidth deviation due to magnetic field inhomogeneity, etc. [16]. Linewidth is represented as full-width at half-maximum. Thus, quantitative estimation is commonly addressed by solving the non-linear least square problems [17], which are relatively complicated and many introduce large errors when the Signal-to-Noise Ratio (SNR) of the input signal is low [15].

MM signal is constituted of all rapidly relaxing signals that have not decayed to zero at the chosen Echo Time (TE). In this work, we also simulated the MM signal by QMEM with Gaussian functions [13], [18].

The quantification tools, such as LCModel [5], QUEST [19], AQSES [20] and TARQUIN [21] etc., developed a measurement of quantitative estimation by designing more empirical priors, constraint conditions and regularizations. However, the process of solving such a large quantitative object is still complicated. With the simulated data test, LCModel was shown that the fitting bias, i.e., the deviation from true values, increases as SNR decreases [22]. Some optimization algorithms of quantifications were also proposed [23]-[26], but their hyperparameters need to be adjusted to achieve the best performance in different cases.

Recently, deep learning has made significant progress in MRS processing [9], [13], [18], [27]-[36]. Its database learning with labels has directly and successfully quantified metabolite concentrations from input spectrum, especially from low SNR spectrum, in an end-to-end manner [31], [33], [35], [36]. Lee *et al.* [18] proposed to predict a clean metabolite spectrum with an end-to-end network and then quantify the metabolite with Linear Least Squares (LLS). However, the LLS in [18] was not included in network training, thus the network cannot feed back the quantification error into metabolite spectrum estimation. Thus, the end-to-end strategy has limited generalization, and the network should be retrained if any metabolite concentration for the test is out of the range from the trained data [18], [27].

In this work, we designed an MRS Quantification neural Network (QNet) with two modules to extract the IFs and overall MM signal. Rather than using end-to-end deep learning, QNet employs LLS aided by the deep neural network of the two modules to quantify metabolite concentrations. The LLS in the proposed QNet takes part in the backpropagation of network, feedbacking the quantification error to into metabolite spectrum estimation, which greatly improves the generalization. We also applied the QMEM to synthesize the basis-set and incorporate *in vivo* parameter information to form many spectra to train QNet. Both simulation and *in vivo* experiments were conducted to compare QNet with LCModel. The former experiments show that QNet can achieve more accurate quantification and be more robust to noise. The latter experiments demonstrate that the quantification results of QNet and LCModel are consistent when SNR is high, whereas QNet can be more stable when SNR decreases.

The rest of this paper is organized as follows: Sections II, III and IV present the proposed method, network implementation, and prepared data, respectively. Section V provides the main results. Section VI discusses the limitations and Section VII concludes this work.

## II. METHOD

This section describes the proposed method that estimates the non-ideal acquisition factors and overall MM signal with deep learning, and then linearly estimates the metabolic concentration.

A schematic diagram of determining the contributions of individual metabolites to the total MRS signal is shown in Fig. 1. The complex matrix $\mathbf{M}$ is the basis-set including individual metabolites. $\mathbf{M}$ is with the size of 512×17 in our case, where 512 is the length of the spectra and 17 is the number of the metabolites. The complex detected spectrum $\mathbf{y} = \mathbf{Mc} + \boldsymbol{\varepsilon}$ is modeled as the sum of additive Gaussian noise $\boldsymbol{\varepsilon}$ and linear combination with the concentration $\mathbf{c}$ of individual metabolites. Concentration $\mathbf{c}$ could be estimated through solving the LLS problem:

$$\hat{\mathbf{c}} = \arg\min_{\mathbf{c}} \left\| \mathbf{M}_{\text{real}} \mathbf{c} - \mathbf{y}_{\text{real}} \right\|_2^2, \quad (1)$$

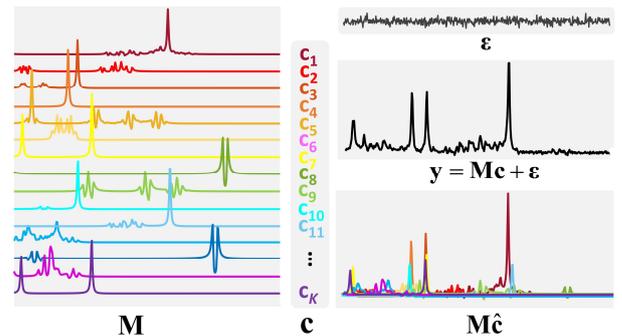

Fig. 1. Schematic diagram of basic spectrum fitting for quantification. Notes: $\mathbf{M}$ is the metabolite basis-set; $\boldsymbol{\varepsilon}$ is the Gaussian noise; $\mathbf{y}$ is the detected spectrum; $\hat{\mathbf{c}}$ is the solution obtained by LLS and thus is an unbiased estimation of concentration. For simplicity, all the parts on the diagram have only real part but they are not shown.

where $\| \cdot \|_2$ is the $l_2$ norm of a vector, $(\cdot)_{\text{real}}$ is taking the real part of $(\cdot)$, and $\hat{\mathbf{c}}$ is the estimated concentration. Equation (1) is effective since it has analytical solution $\hat{\mathbf{c}} = (\mathbf{M}_{\text{real}}^H \mathbf{M}_{\text{real}})^\dagger \mathbf{M}_{\text{real}}^H \mathbf{y}_{\text{real}}$ where $H$ is the Hermitian transpose and $\dagger$ is the pseudo inverse. Besides, the solution $\hat{\mathbf{c}}$ is a minimum-variance unbiased estimation because $\boldsymbol{\varepsilon}$ is homoscedastic and uncorrelated to $\mathbf{Mc}$.

In practice, IFs including phase shift $\Delta\varphi$, frequency shift $\Delta \mathbf{f}$ and linewidth deviation $\Delta d$ should be considered in the quantification. Accordingly, (1) is modified as:

$$\min_{\mathbf{c},\Delta\varphi,\Delta\mathbf{f},\Delta\alpha} \left\| \mathbf{M}'_{\text{real}} \mathbf{c} - \mathbf{y}_{\text{real}} \right\|_2^2, \; \mathbf{M}' = h(\mathbf{M},\Delta\varphi,\Delta\mathbf{f},\Delta d), \quad (2)$$

where the function $h(\mathbf{M},\Delta\varphi,\Delta\mathbf{f},\Delta d)$ is the modulation of $\mathbf{M}$ by IFs $\{\Delta\varphi,\Delta\mathbf{f},\Delta d\}$). In addition to the more difficulty to find the solution, the IFs $\{\Delta\hat{\varphi},\Delta\hat{\mathbf{f}},\Delta\hat{d}\}$ extracted from the input $\mathbf{y}$ by this nonlinear least square estimation of (2) have large errors when the SNR of the input signal $\mathbf{y}$ is low [15].

In the proposed QNet, we designed a neural network module to extract the IFs (upper part of Fig. 2b), and the difficulty of solving (2) is reduced by deep learning:

$$\begin{aligned} &\mathcal{N}_{\text{extraction}}(\mathbf{y}_{\text{real}} \mid \Theta_{\text{extraction}}) = \{\Delta\hat{\varphi},\Delta\hat{\mathbf{f}},\Delta\hat{d}\} \\ &\min_{\mathbf{c}} \left\| h(\mathbf{M},\Delta\hat{\varphi},\Delta\hat{\mathbf{f}},\Delta\hat{d})_{\text{real}} \mathbf{c} - \mathbf{y}_{\text{real}} \right\|_2^2 \end{aligned}, \quad (3)$$

where $\mathcal{N}_{\text{extraction}}(\mathbf{y}_{\text{real}} \mid \Theta_{\text{extraction}})$ is the deep learning model for extracting the IFs from the input $\mathbf{y}_{\text{real}}$ with the trainable network parameters $\Theta_{\text{extraction}}$, and $\{\Delta\hat{\varphi},\Delta\hat{\mathbf{f}},\Delta\hat{d}\}$ is the predicted IFs. Once these IFs have been obtained, estimating $\mathbf{c}$ becomes an easy LLS problem and a closed-form solution can be achieved.

The background signal from MMs should also be considered. So, the parametric model for MMs should be added to solve (2), further exacerbating the difficulty of solving the quantification:

$$\min_{\mathbf{c},\Delta\varphi,\Delta\mathbf{f},\Delta\alpha,\lambda,\Omega} \left\| h(\mathbf{M},\Delta\varphi,\Delta\mathbf{f},\Delta d)_{\text{real}} \mathbf{c} - \mathbf{y}_{\text{real}} + \lambda \mathbf{b}(\Omega) \right\|_2^2, \quad (4)$$

where $\mathbf{b}$ is the background signal from MMs with a structure parameter $\Omega$ and an intensity $\lambda$. Similar to the recent research [33], another deep learning module (lower part of Fig. 2b) is designed in QNet to predict the overall MM signal $\hat{\mathbf{b}}$ which is then subtracted from $\mathbf{y}$. This can reduce the burden of quantitative estimation in (4):

$$\begin{aligned} &\mathcal{N}_{\text{prediction}}(\mathbf{y}_{\text{real}} \mid \Theta_{\text{prediction}}) = \hat{\mathbf{b}} \\ &\min_{\mathbf{c},\Delta\varphi,\Delta\mathbf{f},\Delta\alpha} \left\| h(\mathbf{M},\Delta\varphi,\Delta\mathbf{f},\Delta d)_{\text{real}} \mathbf{c} - \mathbf{y}_{\text{real}} + \hat{\mathbf{b}} \right\|_2^2 \end{aligned}, \quad (5)$$

where $\mathcal{N}_{\text{prediction}}(\mathbf{y}_{\text{real}} \mid \Theta_{\text{prediction}})$ is a deep learning module for predicting $\mathbf{b}$ from the input $\mathbf{y}_{\text{real}}$ with parameters $\Theta_{\text{prediction}}$.

In summary, three steps are required in QNet: 1) Estimating IFs by solving (3); 2) Obtaining the overall MM signal by solving (5). 3) Estimating metabolite concentrations via the LLS problem between the basis-set with IFs and the remaining signal after deduction of the overall MM signal. These steps are expressed as follows:

$$\begin{aligned} &\mathcal{N}_{\text{extraction}}(\mathbf{y}_{\text{real}} \mid \Theta_{\text{extraction}}) = \{\Delta\hat{\varphi},\Delta\hat{\mathbf{f}},\Delta\hat{d}\} \\ &\mathcal{N}_{\text{prediction}}(\mathbf{y}_{\text{real}} \mid \Theta_{\text{prediction}}) = \hat{\mathbf{b}} \\ &\min_{\mathbf{c}} \left\| h(\mathbf{M},\Delta\hat{\varphi},\Delta\hat{\mathbf{f}},\Delta\hat{d})_{\text{real}} \mathbf{c} - \mathbf{y}_{\text{real}} + \hat{\mathbf{b}} \right\|_2^2 \end{aligned}. \quad (6)$$

## III. IMPLEMENTATION OF QNET

Fig. 2 shows the procedure of QNet, which includes: 1) Synthesizing training data; 2) Training two deep learning modules that determine IFs and the overall MM signal from input, respectively; 3) Estimating metabolite concentrations with LLS.

### A. Synthetic Data Generation with QMEM

We resorted to synthetic data to train the network since it is difficult to estimate the true values of the IFs and the concentrations from *in vivo* data with noise, limited resolution and overlap. Synthetic data will be generated following the physical law and this strategy has been observed powerful in deep learning processing of spectroscopy [8], [9], [34]. A complex sampling point of the MRS signal $\mathbf{y}$ is generated according to:

$$\begin{aligned} &\hat{y}(n\Delta t) = \sum_{k=1}^{K} c_k h_k(m_k(n\Delta t) \mid \boldsymbol{\theta}_k) + b(n\Delta t) + \varepsilon(n\Delta t) \\ &n = 0,1,2,..,N-1 \end{aligned}, \quad (7)$$

where $N,\Delta t,K$ denotes the length of the signal, the sampling interval and the number of metabolites, respectively; $c_k$ is the $k^{\text{th}}$ concentration; $\varepsilon(n\Delta t)$ is the complex white Gaussian noise [37]; $b(n\Delta t)$ represents the complex total signal from MMs; and $h_k(m_k(n\Delta t) \mid \boldsymbol{\theta}_k)$ denotes the $k^{\text{th}}$ metabolite component signal modulated with the IFs $\boldsymbol{\theta}_k = \{\Delta\varphi,\Delta\mathbf{f}_k,\Delta d\}$, and is expressed as:

$$\begin{aligned} &h_k(m_k(n\Delta t) \mid \boldsymbol{\theta}_k) = e^{i\Delta\varphi} m_k(n\Delta t) e^{((i2\pi\Delta f_k - \Delta d)n\Delta t)} \\ &n = 0,1,2,...,N-1 \end{aligned}, \quad (8)$$

where $i$ denotes the imaginary unit and satisfies $i^2 = -1$, $\boldsymbol{\theta}_k$ is the set of the IFs (including zero-order phase drift $\Delta\varphi$, frequency drift $\Delta f_k$ and linewidth deviation $\Delta d$) of the $k^{\text{th}}$ metabolite, and $m_k(n\Delta t)$ denotes the $n^{\text{th}}$ point of ideal MRS time-domain signal of the $k^{\text{th}}$ metabolite. The corresponding frequency-domain of $m_k(n\Delta t)$ is the point of the $n^{\text{th}}$ row and $k^{\text{th}}$ column of the basis set matrix $\mathbf{M}$, which is generated by FID-A based on QMEM [11].

### B. Extraction of Imperfection Factors

This module consists of 3 Stacked Convolutional Blocks (SCBs) and 2 Fully Connected Layers (FCLs). The 3 SCBs all contain 2 convolutional layers and use respectively 16, 32 and 64 filters with the same kernel size of $3 \times 1$. After the convolution operation, the non-linear activation function Rectified Linear Unit (ReLU) follows, and then the maximum pooling is performed. The module finally outputs a 1D vector containing a set of IFs for all the target metabolites. The structure of this module is shown in the upper part of Fig. 2b, and the detailed structures of an SCB and an FCL are shown in the middle of Fig. 2b. The overall process of the IF extraction module can be expressed as:

$$\mathcal{N}_{\text{extraction}}(\mathbf{y}_{\text{real}} \mid \Theta_{\text{extraction}}) = \{\Delta\hat{\varphi},\Delta\hat{\mathbf{f}},\Delta\hat{d}\}, \quad (9)$$

where $\Theta_{\text{extraction}}$ represents the network parameters including 3 SCBs and 2 FCLs, and $\mathcal{N}_{\text{extraction}}(\mathbf{y}_{\text{real}} \mid \Theta_{\text{extraction}})$ represents the nonlinear extraction mapping from the detected spectrum $\mathbf{y}_{real}$ to the IFs $\{\Delta\hat{\varphi},\Delta\hat{\mathbf{f}},\Delta\hat{d}\}$.

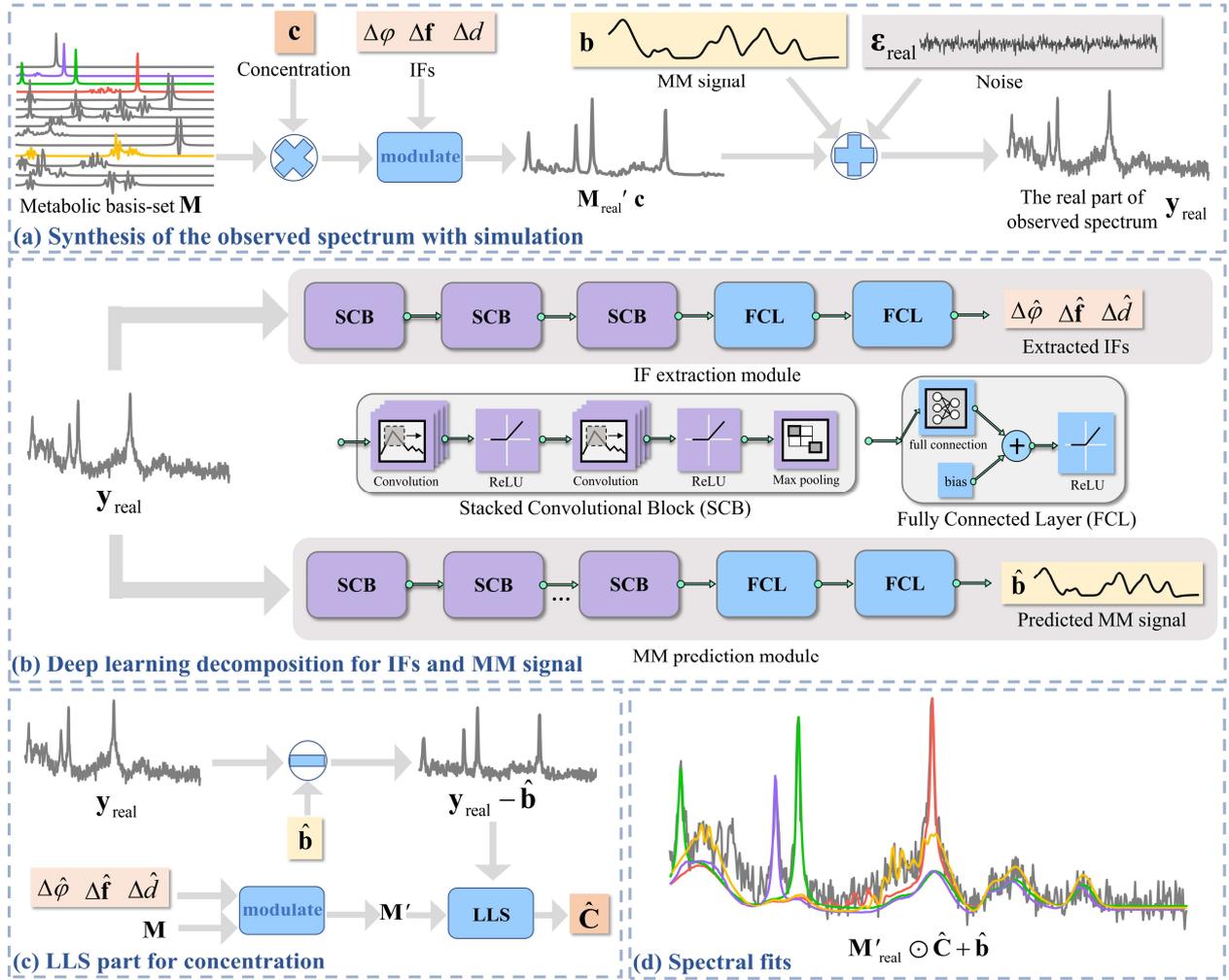

Fig. 2. The procedure of QNet: (a) Procedure of spectrum synthesis, (b) deep learning part with two modules to extract IFs and the overall MM signal, respectively, (c) LLS part using the estimate results from (b) to predict metabolite concentration, and (d) the spectral fits estimated by the real part of the modulated basis set $\hat{\mathbf{M}}(=\mathbf{M}'\odot\hat{\mathbf{C}})$ and the predicted $\hat{\mathbf{b}}$. $\hat{\mathbf{M}}$ is one of the QNet final outputs and modulated by the predicted IFs from the network in (b) and the concentrations estimated by LLS from (c).

### C. Estimation of Overall Macromolecule Signal

This module consists of 6 SCBs and 2 FCLs. The 6 SCBs all consist of 2 convolutional layers, and use 16, 32, 64, 128 and 256 filters, respectively, with the same kernel size of $3\times1$. After the convolution operation, the ReLU activation function follows, and then the maximum pooling is performed. The output of this module is a 1D vector of a size $512\times1$ as predicted MM signal. The structure of this module is shown in the lower part of Fig. 2b. The overall process of the MM prediction module can be expressed as:

$$\mathcal{N}_{prediction}(\mathbf{y}_{real}\mid\Theta_{prediction})=\hat{\mathbf{b}}, \qquad (10)$$

where $\Theta_{prediction}$ represents the network parameters of this module, and $\mathcal{N}_{prediction}(\mathbf{y}_{real}\mid\Theta_{prediction})$ represents the non-linear prediction mapping from the detected spectrum $\mathbf{y}_{real}$ to the background MM signal $\hat{\mathbf{b}}$.

### D. Estimation of Metabolite Concentrations

The LLS part (Fig. 2c) is designed to predict the metabolite concentrations, which are linear parameters in (1). Before the operation of LLS, the basis-set $\mathbf{M}$ is updated to $\mathbf{M}'$ via (8) with $\{\Delta\hat{\varphi},\Delta\hat{\mathbf{f}},\Delta\hat{d}\}$ that are determined by the IF extraction module, and the predicted MM signal $\hat{\mathbf{b}}$, which is subtracted from the detected signal $\mathbf{y}_{real}$. Therefore, the predicted concentration $\hat{\mathbf{c}}$ can be obtained by LLS:

$$\hat{\mathbf{c}}=(\mathbf{M}'^{H}_{real}\mathbf{M}'_{real})^{\dagger}\mathbf{M}'^{H}_{real}(\mathbf{y}_{real}-\hat{\mathbf{b}}). \qquad (11)$$

As a result, the spectral fitting matrix $\hat{\mathbf{M}}$ containing all the individual metabolites is as:

$$\hat{\mathbf{M}}=\mathbf{M}'\odot\hat{\mathbf{C}}, \qquad (12)$$

where $\odot$ represents the Hadamard product, $\hat{\mathbf{C}}=\begin{bmatrix}\hat{\mathbf{c}}^T\\ \hat{\mathbf{c}}^T\\ \cdots\end{bmatrix}_{N\times K}$ and $T$ is the transpose operation. The LLS module is expressed as:

$$\{\hat{\mathbf{c}}, \hat{\mathbf{M}}\} = \mathcal{N}_R(\{\Delta\hat{\varphi}, \Delta\hat{\mathbf{f}}, \Delta\hat{d}\}, \mathbf{M}, \mathbf{y}, \hat{\mathbf{b}}). \quad (13)$$

The loss function is defined as:

$$Loss(\Theta) = \frac{1}{K}\|\hat{\mathbf{M}} - \mathbf{M}^{label}\|_F^2 + \alpha\|\hat{\mathbf{b}} - \mathbf{b}^{label}\|_F^2, \quad (14)$$

where $\Theta = \{\Theta_{extraction}, \Theta_{prediction}\}$ represents the set of overall network parameters, $\|\ \|_F$ is the Frobenius norm for a matrix, $\mathbf{M}^{label}$ and $\mathbf{b}^{label}$ are the training labels, $\alpha$ is a weight to balance the two loss constraints, $\hat{\mathbf{C}}$ is estimated by LLS with respect to $\mathbf{M}'$ modulated by the predicted IFs, and the predicted overall MM signal $\hat{\mathbf{b}}$ (Fig. 2(c)), and $\hat{\mathbf{M}}$ is the complex basis set modulated by the predicted IFs and $\hat{\mathbf{C}}$.

## IV. DATA PREPARATION

### A. Parameter Ranges for Synthetic Data

Synthetic data were generated according to the prior ranges of metabolite signals, MM signals and IFs [18], [38], [39]. Details are listed below.

The number of metabolites, $K$ in (7) is set to include the common brain metabolites. Their ranges of concentrations (mM) are set according to the literature [18], [38], [39] and listed here: N-acetylaspartate (NAA) (7.5-17), glutamate (Glu) (6.0-12.5), creatine (Cr) (4.5-10.5), Myo-Inositol (mI) (4.0-9.0), glutamine (Gln) (3.0-6.0), taurine (Tau) (2.0-6.0). phosphocreatine (PCr) (3.0-5.5), glutathione (GSH) (1.5-3.0), aspartate (Asp) (1.0-2.0), γ-aminobutyric acid (GABA) (1.0-2.0), glucose (Glc) (1.0-2.0), N-acetylaspartylglutamate (NAAG) (0.5-2.5), glycerophosphocholine (GPC) (0.5-2.0), phosphocholine (PCh) (0.5-2.0), alanine (Ala) (0.1-1.5), Lactate (Lac) (0.2-1.0), and scyllo-inositol (Scyllo) (0.3-0.6).

The overall MM signal is a synthesis of 17 MM signals, and each of them is modelled as a Gaussian function that has three parameters (chemical shift (ppm), amplitude (normalized to the largest last one), linewidth (Hz)). These parameters are (0.90, 0.72, 21), (1.20, 0.28, 19), (1.36, 0.38, 16), (1.63, 0.05, 8), (1.68, 0.05, 13), (1.81, 0.05, 13), (2.02, 0.78, 29), (2.08, 0.78, 21), (2.25, 0.78, 18), (2.97, 0.78, 5), (2.97, 0.3, 14), (3.11, 0.11, 18), (3.22, 0.11, 10), (3.27 0.11, 10), (3.67, 0.71, 34), (3.80, 0.71, 12) and (3.96, 1, 37). To improve data generalization, their amplitudes and linewidths are randomly varied within ±10% and ±20%, respectively. The ratio of the intensity of overall MM to overall metabolite signal is 0.15:1 and randomly varied within ±25%. All the above parameters are referred to the literature [18], [40]-[42].

The ranges of the three IFs $\theta_k = \{\Delta\varphi, \Delta\mathbf{f}_k, \Delta d\}$ are [-5.00, 5.00] degree, [-0.08, 0.08] ppm, and [0.04, 0.12] ppm, respectively, all refer to the previous study [18].

Noise level is expressed by $SNR = Amp_{NAA}/\sigma_{noise}$, where $Amp_{NAA}$ is the amplitude of the methyl peak of NAA at ~2.0 ppm, and $\sigma_{noise}$ is the Standard Deviation (SD) of the spectral noise of 8.0-10.0 ppm. The SNR range is set as [3, 80] to cover that of 8 averages to 128 averages for *in vivo* data. A lower SNR means that the metabolite spectra (also called components) are contaminated more seriously by noise (first row of Fig. 3).

We simulated 60000 sets of MRS data for training and 700 for the test. Each set is amplitude normalized and cropped into 512 points in the range of 0.2 to 4.0 ppm.

### B. In Vivo Data

All *in vivo* data were measured from the Philips 3T scanner with a 32-channel head coil. The fast spin-echo sequence was adopted to acquire $T_1$-weighted brain images for voxel selection, and a point-resolved spectroscopy sequence [43] was used to acquire MRS data with an excitation module for water suppression [44]. Experimental parameters are: TR/TE= 2000 ms/ 35 ms, voxel size = 20 mm × 20 mm × 20 mm, number of data points= 2048, spectral bandwidth= 2000 Hz, Number of Signal Averages (NSA) = 128.

A total of 20 spectra were collected from 20 healthy subjects, respectively, in the following brain regions: 15 spectra from anterior or posterior frontal lobes, 1 from the parietal lobe, 1 from the occipital lobe and 3 from anterior or posterior cingulate gyrus. The human data acquisition was approved by the ethics committee of Xiamen University under Application No. XDYX202206K11.

## V. RESULTS

To evaluate the performance, the proposed QNet is compared with other quantification methods, including the LCModel and two state-of-the-art deep learning methods.

The quantification result of a target metabolite is represented by a relative concentration that is the ratio of the concentration of this metabolite to the concentration of the tCr (Cr + PCr).

### A. Comparison with LCModel with Simulated Data

Fig. 3 shows fitted spectra under representative high (first 2 columns) and low (last 2 columns) SNRs. When the SNR is high (SNR= 71.8), both QNet and LCModel can achieve accurate fits visually since the residual of both methods are small. Besides, QNet obtains a lower residual, which indicates a better fit, than LCModel. When the SNR is low (SNR= 20.0), both methods lead to strong residual and it is hard to judge which method is better.

To quantitatively compare the two methods, 100 sets of simulation data for each of the 7 SNR levels, that is 700 sets in total, were generated for the tests in Fig. 4. The estimated concentration of a component is directly compared with its ground truth value. Metabolites are listed in the ascending order (from left to right) of their true concentrations in the simulated data. The Mean Absolute Percentage Error (MAPE) is calculated as a criterion to evaluate quantification error:

$$MAPE = \frac{1}{I}\sum_{i=1}^{I}\frac{|c_i - \hat{c}_i|}{c_i}, \quad (15)$$

where $I$ is the number of the data set with the same SNR level ($I$ = 100 in this case), $c_i$ is the ground truth value of the *i*-th set of data and $\hat{c}_i$ is the estimation of $c_i$. A lower MAPE means more accurate quantification.

Fig. 4 shows that, for most metabolites, except for the too-low concentration metabolites (Ala, Asp, Lac, GABA), the MAPEs are smaller with QNet than with LCModel. In particular, for NAAG, GPC, PCh and Gln, while LCModel has large MAPEs close to 100%, QNet has significant improvement. Thus, QNet is a good quantification tool, and can even provide more accurate quantifications for many metabolites than LCModel.

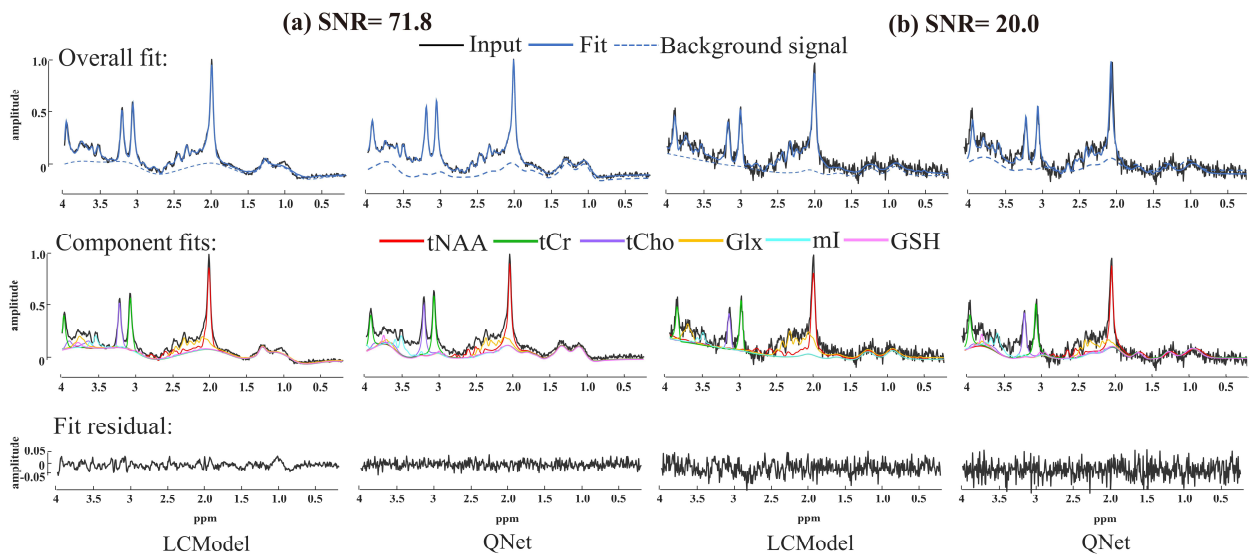

Fig. 3. Spectral fits by LCModel and QNet with the simulation data representative of high and low SNRs of (a) 71.8 and (b) 20.0, respectively. The first row are the original input spectra (black solid line), the overall fit (blue solid line), and the estimated MM signal (blue dash line). The second row are the estimated individual spectra of the 6 main metabolites (tNAA, tCr, tCho, Glx, mI and GSH). The third row are the fitting residuals.

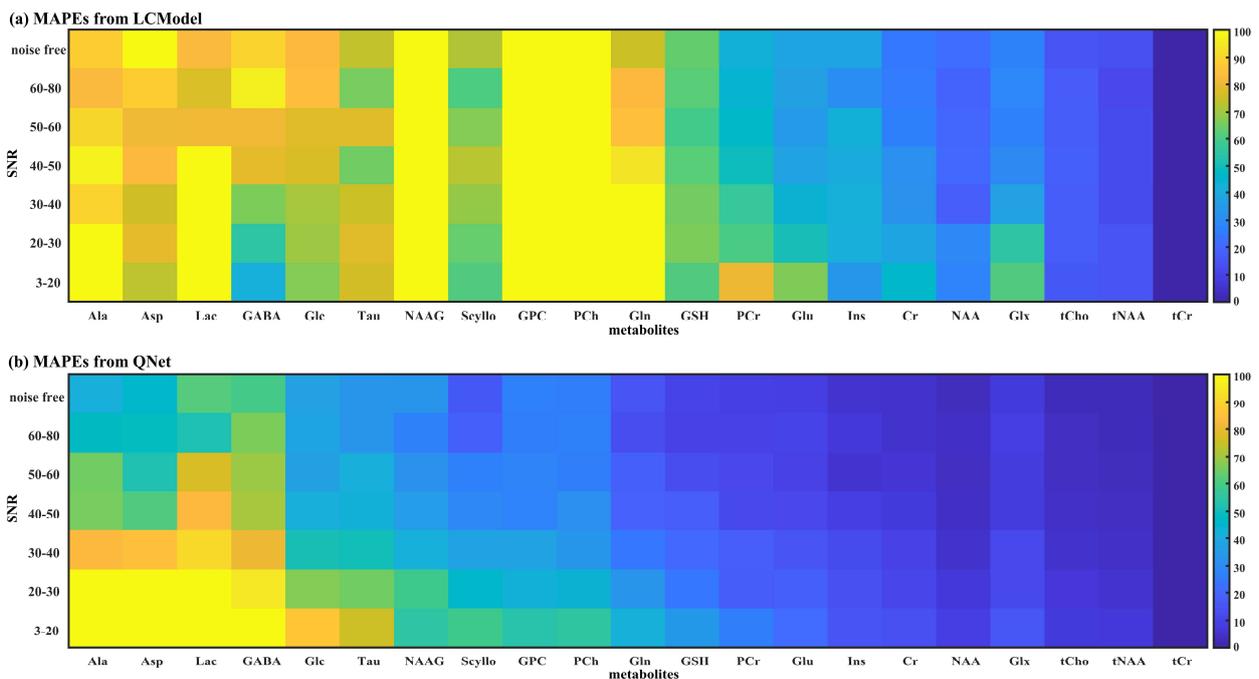

Fig. 4. The MAPEs of the estimated metabolite concentrations by (a) LCModel and (b) QNet at different SNRs. From left to right, metabolites are listed in the ascending order of their true concentrations in simulated data.

### B. Comparison with LCModel with In Vivo Data

Experiments on *in vivo* brain regions were also conducted to compare QNet to LCModel qualitatively in Fig. 5 and quantitatively in Figs. 6-7 and Table I.

Fig. 5 shows the spectrum from an anterior frontal lobe with NSA=128 (high SNR). Both LCModel and QNet have visually similar fitting results with the overall fits, their residuals and the estimations of the components (tNAA, tCr, tCho, Glx, mI and GSH). Both methods have fits of high quality with small residuals.

The consistency of estimated metabolite concentration between LCModel and QNet is further assessed quantitively with the Bland-Altman analysis using 20 *in vivo* spectra with NSA=128 (Fig. 6). First, one-sample T-test is performed with Standard Normal Distribution (SND) and the result of all *p* values > 0.05 is obtained, which means that the distribution of the 20 samples' differences is similar to SND and the systematic bias between the two methods is small. Second, because of $p > 0.05$ (which means normality assumption is satisfied), ± 1.96 SD can be used as the limits of agreement (dash line) to demonstrate a 95% confidence interval of the distributed data. Result shows that 95% or 90% of the data points are within the limits of agreement, which means good consistency between QNet and LCModel. Furthermore, we compute the correlations between the concentrations from LCModel and the proposed method. The Pearson correlation coefficients are 0.90 for tNAA, 0.95 for tCho, 0.61 for mI, 0.50 for GSH and 0.18 for Glx. Thus, mI and GSH (0.4-0.7) have a moderate linear correlation; tNAA and tCho ($\geq 0.9$) have a strong linear correlation; Glx has a low linear correlation.

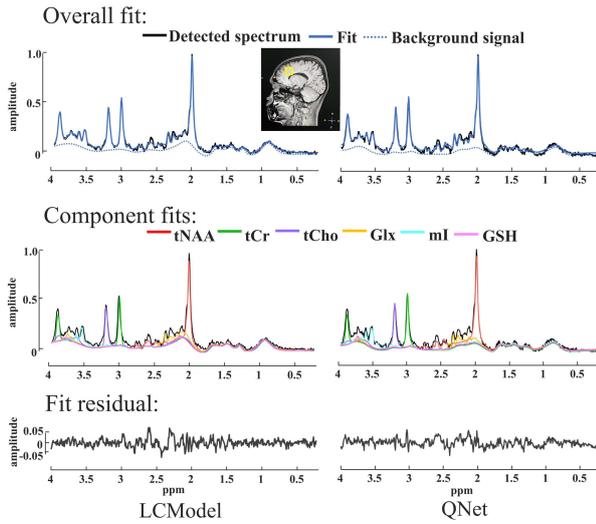

Fig. 5. Comparison between LCModel and QNet with the *in vivo* spectrum from an anterior frontal lobe (in the yellow box). The first row are the original observed spectra (black solid line), the overall fit (blue solid line), and the estimated background signal (blue dash line). The second row are the contributions of the six highest components. The third row are the residuals.

Fig. 7 compares the metabolite concentrations with boxplots estimated by QNet and LCModel from the 20 *in vivo* brain spectra. For tNAA, mI and Glx, both methods can acquire the estimated concentrations all within the normal ranges from [18], [38], [39]. For tCho, a few estimated concentrations reside outside of the upper limit with both methods. As for GSH, both methods get the worst results. Most of the results are outsides of the upper limit for LCModel, while QNet has better quantification with most of the values still within the limits.

Fig. 8 compares the metabolite quantification with different NSAs to imply the performance at different levels of noise for a *in vivo* spectrum. Higher NSA means higher SNR since more averages are used at the cost of longer data acquisition time [27]. The SD is employed to evaluate the robustness of the two methods to different NSAs. Compared with LCModel, the concentrations estimated by QNet are generally more stable with smaller SDs. For Glx, the difference between the SDs of the QNet (2.33) and LCModel (22.63) is very big, and the difference for GSH is big as well. Furthermore, the SDs of all the 20 *in vivo* MRS data with different NSA are listed in Table

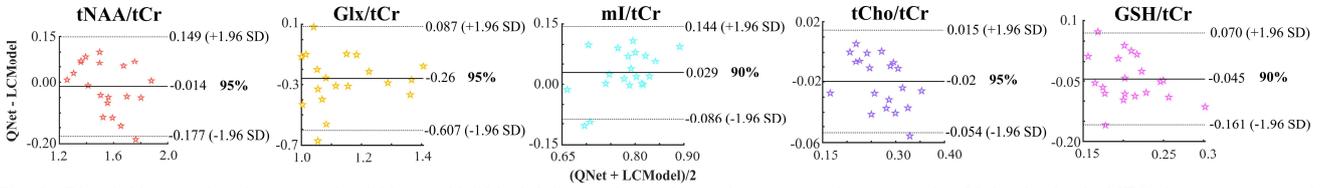

Fig. 6. Bland-Altman plots between the QNet and LCModel fits for the estimated concentrations from the 20 *in vivo* brain MRS data represented as 20 stars, respectively. The difference of the two measurements is plotted on the y axis, and the corresponding average on the x axis. The mean of the differences (solid line) is showed, and the limits of agreement (dashed lines) are calculated as mean ± 1.96 SD. The percentage means the percent of the data points reside within the limits.

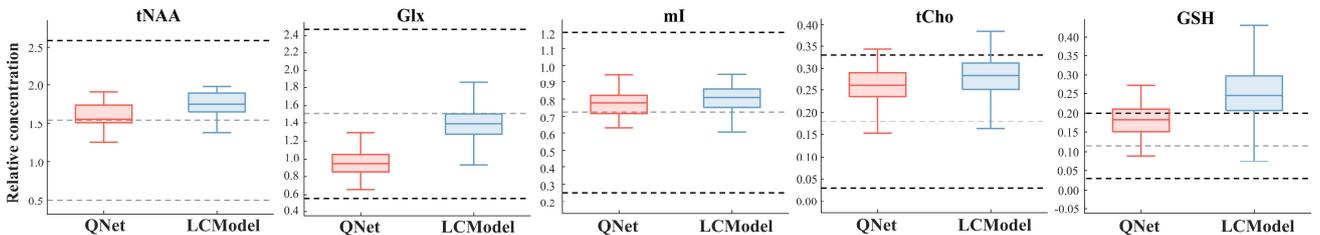

Fig. 7. Comparison of the relative concentrations estimated by QNet (red, left) and LCModel (blue, right) with box plots from the 20 *in vivo* brain spectra. Each box-plot contains 100×20 sets of data, specifically, 100 spectra with NSA of 124 are extracted randomly from each of the 20 fully sampled *in vivo* spectra with NSA of 128. The metabolite concentration range of normal human brain from the literature is marked with the dark dash lines and the mean value is marked with the light dash line.

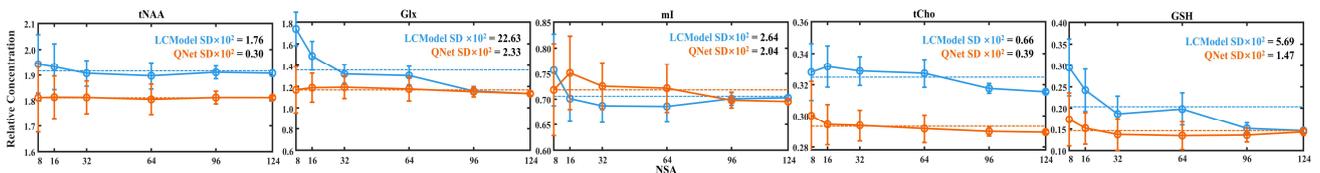

Fig. 8. Comparison of the relative concentrations estimated with different NSA by QNet (red) and LCModel (blue). The 6 data points are acquired from the same fully sampled *in vivo* spectrum of sample 1 with NSA = 8, 16, 32, 64 and 124, respectively.

TABLE I
100× SD OF *IN VIVO* SPECTRA WITH NSA = 8, 16, 32, 64, 96 AND 124

| Sample No. | tNAA/tCr | | mI/tCr | | Glx/tCr | | tCho/tCr | | GSH/tCr | |
|---|---|---|---|---|---|---|---|---|---|---|
| | LCModel | QNet | LCModel | QNet | LCModel | QNet | LCModel | QNet | LCModel | QNet |
| 1 (Fig. 8) | 1.76 | **0.30** | 2.64 | **2.04** | 22.63 | **2.33** | 0.66 | **0.39** | 5.69 | **1.47** |
| 2 | 3.80 | **3.78** | 6.29 | **2.50** | 9.32 | **1.68** | 1.58 | **0.21** | 2.00 | **0.20** |
| 3 | 3.47 | **1.37** | 2.94 | **2.01** | 15.40 | **0.64** | 1.30 | **0.08** | 3.86 | **1.39** |
| 4 | 2.61 | **2.21** | 2.96 | **2.01** | 18.19 | **4.78** | 0.65 | **0.05** | 2.75 | **0.67** |
| 5 | **1.58** | 3.29 | 3.38 | **2.45** | 25.07 | **6.43** | 0.34 | **0.15** | 2.34 | **1.51** |
| 6 | 1.95 | **1.30** | 4.18 | **2.15** | 15.72 | **6.57** | 1.16 | **0.58** | 2.04 | **1.04** |
| 7 | 5.89 | **1.51** | 3.23 | **1.89** | 7.66 | **1.35** | 1.70 | **0.14** | 3.98 | **0.80** |
| 8 | 3.01 | **0.86** | 8.57 | **1.66** | 19.70 | **1.97** | 0.74 | **0.12** | 3.36 | **0.83** |
| 9 | **0.53** | 1.37 | 5.46 | **2.36** | 9.70 | **1.13** | 1.04 | **0.31** | 2.70 | **1.28** |
| 10 | 1.30 | **1.42** | **1.26** | 3.72 | 12.31 | **0.59** | 0.92 | **0.07** | 2.13 | **0.25** |
| 11 | **2.45** | 2.50 | 9.92 | **4.01** | 11.51 | **5.45** | 1.82 | **0.23** | **1.26** | 1.50 |
| 12 | 14.21 | **0.96** | **1.95** | 2.04 | 11.38 | **4.35** | 0.55 | **0.29** | 1.86 | **0.75** |
| 13 | 1.65 | **0.61** | 2.33 | **1.72** | 12.73 | **3.87** | 1.14 | **0.20** | 4.16 | **1.39** |
| 14 | 2.27 | **0.63** | 10.39 | **3.06** | 9.62 | **2.08** | 1.49 | **0.27** | 2.84 | **1.44** |
| 15 | 4.92 | **1.30** | **1.36** | 2.46 | 15.18 | **1.52** | 0.77 | **0.22** | 2.41 | **1.86** |
| 16 | **0.92** | 1.28 | 2.28 | **1.56** | 11.79 | **2.16** | 0.61 | **0.23** | 2.84 | **1.81** |
| 17 | 1.98 | 1.98 | 2.16 | **0.66** | 12.04 | **3.30** | 0.80 | **0.54** | **0.60** | 2.00 |
| 18 | 3.79 | **1.70** | **1.68** | 2.05 | 6.73 | **3.96** | 0.80 | **0.21** | 0.87 | **0.84** |
| 19 | 1.92 | **0.75** | 2.91 | **1.91** | 19.99 | **5.57** | 1.33 | **0.10** | **0.54** | 0.84 |
| 20 | 2.44 | **1.22** | 1.48 | **1.18** | 9.33 | **3.77** | 0.96 | **0.14** | 2.28 | **1.67** |
| QNet score | 16/20 | | 17/20 | | 20/20 | | 19/20 | | 17/20 | |

Note: Lower SDs are highlighted in bold; the score means that the number of times QNet has smaller SDs than LCModel over 20 (the number of spectra).

TABLE II
COMPARISONS OF QNET WITH THE TWO DEEP LEARNING QUANTIFICATION METHODS.

| Method | SNR definition | Metabolite | SNR range of simulated data | SMAPE or MAPE | |
|---|---|---|---|---|---|
| | | | | The compared method | QNet |
| From [31] (10⁴ data) | $\text{SNR}_{\text{FID}} = \frac{[ifft(\mathbf{y})]_{\text{first}}}{\sigma'_{\text{noise}}}$ | NAA | $\text{SNR}_{\text{FID}} = 10$ | 18.95 | **1.98** |
| | | NAAG | | 16.03 | **13.98** |
| | | Cr | | 19.64 | **2.98** |
| | | PCr | | 19.39 | **5.02** |
| | | GPC | | 13.67 | **11.86** |
| | | PCh | | 14.27 | **12.05** |
| | | Glu | | 22.29 | **4.76** |
| | | Gln | | 22.98 | **6.87** |
| | | mI | | 20.20 | **2.84** |
| From [18] (5000 data) | $\text{SNR}_x = \frac{\max(\mathbf{y}[x_k^1, x_k^2])}{\sigma_{\text{noise}}}$ | tNAA | [6.90, 20.74] | **4.00% ± 3.38%** | 4.35% ±2.76% |
| | | tCr | [6.84, 18.26] | 3.66% ± 3.05% | **2.41% ± 1.43%** |
| | | tCho | [7.05, 15.19] | 7.87% ± 7.35% | **2.59% ± 1.57%** |
| | | Glx | [5.28, 9.48] | 5.36% ± 4.48% | **4.56% ± 2.73%** |
| | | mI | [6.23, 13.32] | 7.61% ± 6.77% | **7.47% ± 4.81%** |

Note: SMAPE/MAPE is used for the comparison with the method in [31]/[18], respectively. Lower errors are highlighted in bold.

I, which means that QNet achieves smaller SDs for most of the MRS data than LCModel. Specifically, out of the 20 MRS data, QNet has 16, 17, 20, 19 and 17 smaller SDs for tNAA, mI, Glx, tCho and GSH, respectively. Therefore, QNet is more robust to different NSAs than LCModel.

### C. Comparison with Other Deep Learning Methods

Table II compares QNet with two deep learning methods with simulation data. To make the comparison as fair as possible, the definitions of SNR and forecasting performance evaluation criteria from the reference are adopted. In [31], SNR is defined as $\text{SNR}_{\text{FID}} = [ifft(\mathbf{y})]_{\text{first}}/\sigma'_{\text{noise}}$, where $[ifft(\mathbf{y})]_{\text{first}}$ is the first point of time domain signal of spectrum $\mathbf{y}$, and $\sigma'_{\text{noise}}$ is the SD of the noise in the time domain; and the Symmetric MAPE (SMAPE) [31] is used to measure the accuracy of the model for each metabolite and $\text{SMAPE} = \frac{1}{I}\sum_{i=1}^{I}\frac{|c_i - \hat{c}_i|}{(c_i + \hat{c}_i)}$. In [18], SNR is defined as $\text{SNR}_x = \max(\mathbf{y}[x_k^1, x_k^2])/\sigma_{\text{noise}}$, where $x$ is the target metabolite set {tNAA,tCr,tCho,mI,Glx} and $[x_k^1, x_k^2]$ is the chemical shift range of $k^{\text{th}}$ metabolite target peaks, and $\mathbf{y}[x_k^1, x_k^2]$ is the observed spectrum truncated within the chemical shift range.

Compared with the method from [31], QNet has a smaller SMAPE for all 9 metabolites listed in Table II. While compared with the method from [18], QNet has MAPEs smaller for tCho, Glx, tCr and mI, but slightly larger for tNAA. The comparisons may indicate QNet outperforms the methods from [31] and [18].

### D. Comparison with End-to-End QNet

The generalization ability of deep learning methods with LLS or the end-to-end manner is tested here. Only the range of NAA for the test data was changed to an abnormal range of 0.5-2.0 mM. An end-to-end QNet, which has another module of 6 SCBs with the same settings as the MM prediction module, was built to estimate metabolite concentrations directly from the network. Fig. 9 shows that when a metabolite concentration is not within the range set for trained data, the end-to-end manner cannot fit well and has a large fitting residual, while the QNet with LLS can still have an accurate fit with a small residual.

TABLE III.
THE MAPEs OF QUANTIFICATION OF THE SPECTRA WITH ABNORMAL CONCENTRATIONS BY QNET (CONTAINING LLS) AND END-TO-END QNET (NOT CONTAINING LLS), RESPECTIVELY.

| Method | tNAA | tCr | tCho | Glx | GSH | mI |
|---|---|---|---|---|---|---|
| QNet | **68.3±40.6** | **54.1±38.4** | **25.4±14.2** | 33.6±18.9 | 48.6±32.0 | **17.5±15.6** |
| End-to-end QNet | >516.6 | >544.1 | 42.1±10.4 | **16.0±9.2** | **17.8±11.2** | 19.5±11.4 |

Note: MAPE is shown with mean ±SD.

To further illustrate that the role of the embedded LLS part of QNet in achieving more credible results, according to the literature [45] about metabolite concentration abnormalities in the brains with tumors, we simulated 100 such high SNR (~80) spectra with the metabolite concentrations that are out of the ranges for training. Specifically, tNAA is abnormally lowered to 0.5-2.0 mM from healthy 7.5-17.5 mM, tCho is abnormally higher to 5.0-12.0 mM from healthy 1.0-4.0 mM and tCr is abnormally lowered to 0.5-3.0 mM from healthy 7.5-16.0 mM. As can be seen in Table III, for the challenging quantification of the abnormally low tNAA and tCr concentrations, the errors of QNet are the smaller than those of end-to-end QNet, which indicates that the QNet is much more credible than end-to-end QNet. For the abnormally high tCho, QNet also has better performance than the end-to-end QNet. However, since LLS is an estimation of the overall spectrum, abnormalities in tNAA, tCho, and tCr cause bigger errors compared to the normal situations (MAPE<20), which may affect the quantification of other metabolites. Therefore, QNet has slightly higher quantification errors of Glx and GSH with normal ranges than the end-to-end QNet.

To demonstrate that QNet with LLS also has better generalization for varying SNRs in *in vivo* situation, we tested the two versions of QNet with 20 *in vivo* brain spectra with different NSAs. In terms of consistency and rationality of high SNR situation (NSA=124), there is no significant difference between the QNet with LLS and end-to-end QNet (Figure S1-2 in supplementary materials). But for the stability (Table IV) of varying NSAs, QNet reflects a much smaller variance in quantification of the five main metabolites.

## VI. DISCUSSIONS

### A. Hyper-parameter settings of QNet

An ablation study is designed to justify our two neural networks of the IF extraction module and MM prediction module. Specifically, we trained the following two models, one is the QNet without the IF extraction module (called QNet-woIF) and the other one is the QNet without the MM prediction module (called QNet-woMM). Fig. S3 in supplementary materials shows that, without one of the two modules, QNet will lead to much higher quantification errors which are greater than 100%. Therefore, both modules are indispensable to our QNet.

Lineshape is an important factor for quantifying the experimental data while it is often difficult to judge whether a peak in a spectrum is Lorentzian, Gaussian or Voigt [46]. QNet is considered the global Lorentzian lineshape and we also trained QNet-Voigt which is the QNet trained by Voigt-lineshape synthetic data. The number of nonlinear parameters to be estimated greatly increases from 19 (1 global phase shift, 1 global Lorentzian linewidth and 17 metabolite dependent frequency shifts) to 52 (1 global phase shift, 17 metabolite dependent Lorentzian linewidth components, 17 metabolite dependent Gaussian linewidth components and 17 metabolite dependent frequency shifts), which makes QNet-Voigt much more complex than QNet. Table S1 in supplementary materials shows that for *in vivo* test, QNet-Voigt is less stable than QNet, when NSA changes. The proposed QNet achieves better performance by considering only the global Lorentzian lineshape.

From the Fourier theory, FID and its spectrum have the equivalent information. We chose spectra as input because of the successful cases of previous studies [18], [31], [33], [35]. All of them showed that applying convolutional neural network can extract the features of spectra. Nevertheless, it is also feasible to use the time domain FIDs as input [36]. For the proposed QNet, there is no significant difference of quantification errors between the two kinds of inputs with complex spectra and only the real part of spectra (Figure S4 in supplementary materials). A possible reason is that our nonlinear parameter estimation module of the proposed QNet is an end-to-end nonlinear mapping which is task-oriented and not sensitive to imaginary part of input.

### B. Limitations on QNet

Physics-driven synthetic data used for training is an emerging method, which has been evidenced convincing in recent review papers [8], [9]. It effectively alleviates the situation that there are not enough real data.

To test the reliability of QNet which is trained by physics-driven synthetic data, we set four types of misspecifications of the noise level, ranges of metabolite concentrations, ranges of IFs and *in vivo* situation. For the misspecification of the noise level, the SNR ranges are set out of the range for training, and QNet can obtain a high accuracy when the noise is very low (Table S2 in supplementary materials), but the performance of QNet is seriously affected by strong noise when metabolite signal is weak (Table S3). For the misspecification of the concentrations, as can be seen in Table III, QNet has better generalization, when quantifying the synthetic spectrum with the metabolite concentrations out of the ranges for training. However, since LLS is an estimation of the overall spectrum, one or more metabolic abnormalities may affect the quantification of other metabolites with normal ranges. For the misspecification of ranges of IFs, we extended the ranges of IFs for test. Table S4 shows that QNet has acceptable quantification errors. The *in vivo* data have natural misspecification with the synthetic training data. Even so, QNet reflects a much smaller variance in quantification of the five main metabolites (shown in Table IV).

TABLE IV.
100×SD OF *In Vivo* Spectra with Different NSA (8, 16, 32, 64, 96 AND 124)

|   | tNAA/tCr | | Glx/tCr | | mI/tCr | | tCho/tCr | | GSH/tCr | |
|---|---|---|---|---|---|---|---|---|---|---|
|   | End-to-end QNet | QNet | End-to-end QNet | QNet | End-to-end QNet | QNet | End-to-end QNet | QNet | End-to-end QNet | QNet |
| 1 | 6.64 | **0.30** | 36.34 | **2.33** | 10.73 | **2.04** | 2.99 | **0.39** | 5.93 | **1.47** |
| 2 | 11.14 | **3.78** | 13.27 | **1.68** | 7.47 | **2.50** | 1.44 | **0.21** | 6.97 | **0.20** |
| 3 | 8.45 | **1.37** | 10.04 | **0.64** | 4.18 | **2.01** | 1.41 | **0.08** | 6.37 | **1.39** |
| 4 | 2.76 | **2.21** | 6.45 | **4.78** | 5.44 | **2.01** | 0.88 | **0.05** | 3.53 | **0.67** |
| 5 | 4.85 | **3.29** | 9.82 | **6.43** | 4.77 | **2.45** | 2.42 | **0.15** | 2.41 | **1.51** |
| 6 | 4.66 | **1.30** | 16.20 | **6.57** | 4.66 | **2.15** | 1.34 | **0.58** | 4.25 | **1.04** |
| 7 | 5.91 | **1.51** | 7.84 | **1.35** | 21.3 | **1.89** | 0.90 | **0.14** | 5.14 | **0.80** |
| 8 | 7.50 | **0.86** | 17.69 | **1.97** | 7.19 | **1.66** | 1.50 | **0.12** | 5.23 | **0.83** |
| 9 | 1.49 | **1.37** | 10.16 | **1.13** | 4.41 | **2.36** | 0.65 | **0.31** | 4.07 | **1.28** |
| 10 | 3.41 | **1.42** | 4.66 | **0.59** | 4.48 | **3.72** | 0.69 | **0.07** | 2.51 | **0.25** |
| 11 | 6.68 | **2.50** | 25.60 | **5.45** | 20.97 | **4.01** | 1.06 | **0.23** | 3.94 | **1.50** |
| 12 | 23.70 | **0.96** | 18.84 | **4.35** | 14.33 | **2.04** | 3.55 | **0.29** | 12.03 | **0.75** |
| 13 | 13.62 | **0.61** | 5.63 | **3.87** | 2.29 | **1.72** | 0.45 | **0.20** | 4.02 | **1.39** |
| 14 | 5.86 | **0.63** | 6.52 | **2.08** | 16.66 | **3.06** | 1.16 | **0.27** | 4.66 | **1.44** |
| 15 | 6.90 | **1.30** | 13.54 | **1.52** | 2.68 | **2.46** | 0.89 | **0.22** | 1.86 | 1.86 |
| 16 | 8.73 | **1.28** | 5.16 | **2.16** | 2.51 | **1.56** | 1.04 | **0.23** | 3.15 | 1.81 |
| 17 | 3.71 | **1.98** | 10.16 | **3.30** | 17.61 | **0.66** | 1.18 | **0.54** | 4.65 | 2.00 |
| 18 | 4.19 | **1.70** | 9.18 | **3.96** | 5.19 | **2.05** | 0.43 | **0.21** | 4.97 | **0.84** |
| 19 | 9.32 | **0.75** | 32.36 | **5.57** | 12.69 | **1.91** | 1.62 | **0.10** | 2.41 | **0.84** |
| 20 | 4.41 | **1.22** | **2.75** | 3.77 | 2.41 | **1.18** | 1.22 | **0.14** | **1.04** | 1.67 |
| QNet score | 20/20 | | 19/20 | | 20/20 | | 20/20 | | 18/20 | |

Note: Lower SDs are highlighted in bold; the score means that the number of times that QNet has smaller SDs than end-to-end QNet over 20 subjects *in vivo* data.

Still, as an emerging field, large-scale real-data validation is still required to test the model, and we will collaborate with clinics to further verify the reliability in the future.

There are challenging peaks difficult to quantify. It is still not very accurate to quantify some metabolites which are weak or overlapped with other strong resonance signals (e.g., Ala, Asp, Lac, GABA, etc.), and distinguish the metabolites with similar structures (e.g., NAAG and NAA, Cr and PCr, etc.) with QNet. At a clinical magnetic field strength of 3.0 T, it is still difficult to distinguish them from each other by the spectrum obtained with the standard point resolved spectroscopy sequence excitation. Therefore, simulation of the MRS data acquired from specific sequences (e.g., MEGA- point-resolved spectroscopy sequence which is designed specifically to measure GABA) to train QNet is expected to address this difficulty.

Diversity of the training data is limited. Extension of the diversity of the simulated data is necessary. Since the simulated data used in this work are referenced from the literature all with healthy subjects, more data from unhealthy brains need to be collected, and the ranges of metabolite concentrations need to be updated. Or transfer learning can be applied to QNet to obtain a new quantification model suitable for both healthy and unhealthy brain MRS data to promote clinical applications.

Absolute concentration quantification is expected. Only relative concentrations of metabolites are presented. This approach may be limited by the stability of tCr level in subjects. Absolute quantification of metabolites is more challenging but may provide quantitive diagnosis report for patients.

## VII. CONCLUSION

In this work, we proposed the MRS Quantification deep learning Network (QNet). It combines the superior nonlinear learning capability of neural networks and the effectiveness of Linear Least Squares (LLS). The training data is generated as practical as possible through simulating the metabolite basis-set with the physics-informed quantum mechanical and exponential model, and incorporating practical metabolite concentration ranges, imperfection factors, macromolecular background signal and noise. QNet first applied multi-layer convolutions to predict the imperfection factors and the macromolecular background signal separately to simplify the problem of least squares from complicated nonlinearity to simple linearity, and then LLS is applied to linearly estimate individual metabolite concentrations. This special strategy improves the generalization of the common end-to-end deep learning method. Results show that, for the simulated data, QNet can achieve lower quantification errors than LCModel for most (80%) metabolites, especially at low signal-to-noise level; for the healthy *in vivo* data, QNet gets the quantification results that are highly consistent with those of LCModel in the case of high signal-to-noise level, and achieves more stable quantification at low signal-to-noise level. In summary, this study provides an intelligent, reliable and robust MRS quantification method. QNet has been deployed on a cloud computing platform, CloudBrain-MRS, which is open accessed at https://csrc.xmu.edu.cn/CloudBrain.html (test account: DEMO; password: CSG12345678! ).